\input jytex.tex   
\typesize=10pt \magnification=1200 \baselineskip=17truept
\hsize=6truein\vsize=8.5truein

\sectionnum=0
\sectionnumstyle{arabic}
\chapternumstyle{blank}
\chapternum=1
\pagenum=0

\def\begintitle{\pagenumstyle{blank}\parindent=0pt\begin{narrow}[0.4in]}
\def\endtitle{\end{narrow}\newpage\pagenumstyle{arabic}}


\def\beginproblem{\vskip 20truept\parindent=0pt\begin{narrow}[10 truept]}
\def\endproblem{\vskip 10truept\end{narrow}}


\def\eql#1{\eqno\eqnlabel{#1}}

\def\ref{\reference}
\def\peq{\puteqn}
\def\pref{\putref}

\def\mgn{\marginnote}
\def\bpr{\begin{problem}}
\def\epr{\end{problem}}


 
\def\mbox#1{{\leavevmode\hbox{#1}}}
\def\hspace#1{{\phantom{\mbox#1}}}



\def\Ga{\Gamma}

\def\La{\Lambda}

\def\ze{\zeta}
\def\bze{{\bmit\zeta}}

\def\zf{$\zeta$-- function}
\def\zfs{$\zeta$-- functions}


\def\frac#1/#2{\leavevmode\kern.1em
\raise.5ex\hbox{\the\scriptfont0 #1}\kern-.1em/\kern-.15em
\lower.25ex\hbox{\the\scriptfont0 #2}}
\def\sfrac#1/#2{\leavevmode\kern.1em
\raise.5ex\hbox{\the\scriptscriptfont0 #1}\kern-.1em/\kern-.15em
\lower.25ex\hbox{\the\scriptscriptfont0 #2}}

\def\gtorder{\mathrel{\raise.3ex\hbox{$>$}\mkern-14mu
             \lower0.6ex\hbox{$\sim$}}}
\def\ltorder{\mathrel{\raise.3ex\hbox{$<$}|mkern-14mu
             \lower0.6ex\hbox{\sim$}}}

\def\semidirprod{\rlap{\ss C}\raise1pt\hbox{$\mkern.75mu\times$}}
\def\for{\lower6pt\hbox{$\Big|$}}
\def\fish{\kern-.25em{\phantom{abcde}\over \phantom{abcde}}\kern-.25em}

 

\def\boxit#1{\vbox{\hrule\hbox{\vrule\kern3pt
        \vbox{\kern3pt#1\kern3pt}\kern3pt\vrule}\hrule}}
\def\dalemb#1#2{{\vbox{\hrule height .#2pt
        \hbox{\vrule width.#2pt height#1pt \kern#1pt
                \vrule width.#2pt}
        \hrule height.#2pt}}}



\def\eg{{\it e.g. }}

\def\pa{\partial}
  

\def\tr{{\rm tr\,  }}

\def\sumdash#1{{\mathop{{\sum}'}_{#1}}}
\def\sumdash2#1#2{{\mathop{{\sum}'}_{#1}^{#2}}}

\def\3j#1#2#3#4#5#6{\left\lgroup\matrix{#1&#2&#3\cr#4&#5&#6\cr}
\right\rgroup}

\def\man{{\cal M}}
\def\cad{{\cal D}}

\def\m?{\mgn{?}}


\def\beginexercise{\vskip 20truept\parindent=0pt\begin{narrow}[10 truept]}
\def\endexercise{\vskip 10truept\end{narrow}}

\def\beginremark{\vskip 10truept\parindent=0pt\begin{narrow}[10 truept]}
\def\endremark{\vskip 10truept\end{narrow}}

\def\beginexample{\vskip 10truept\parindent=0pt\begin{narrow}[10 truept]}
\def\endexample{\vskip 10truept\end{narrow}}

\def\bex{\begin{exercise}}
\def\eex{\end{exercise}}
\def\ber{\begin{remark}}
\def\eer{\end{remark}}
\def\bexa{\begin{example}}
\def\eexa{\end{example}}


\def\aop#1#2#3{{\it Ann. Phys.} {\bf {#1}} (19{#2}) #3}

\def\cmp#1#2#3{{\it Comm. Math. Phys.} {\bf {#1}} (19{#2}) #3}
\def\cqg#1#2#3{{\it Class. Quant. Grav.} {\bf {#1}} (19{#2}) #3}

\def\jmp#1#2#3{{\it J. Math. Phys.} {\bf {#1}} (19{#2}) #3}
\def\jpa#1#2#3{{\it J. Phys.} {\bf A{#1}} (19{#2}) #3}

\def\np#1#2#3{{\it Nucl. Phys.} {\bf B{#1}} (19{#2}) #3}
\def\pl#1#2#3{{\it Phys. Lett.} {\bf {#1}} (19{#2}) #3}

\def\prD#1#2#3{{\it Phys. Rev.} {\bf D{#1}} (19{#2}) #3}

\def\cras#1#2#3{{\it Comptes Rend. Acad. Sci. (Paris)} {\bf{#1}} (#2) #3}

\def\invm#1#2#3{{\it Invent. Math.} {\bf {#1}} (19{#2}) #3}
\def\jdg#1#2#3{{\it J. Diff. Geom.} {\bf {#1}} (19{#2}) #3}

\def\tams#1#2#3{{\it Trans. Am. Math. Soc.} {\bf {#1}} (19{#2}) #3}

\begin{title}
\vglue 1truein
\vskip15truept
\centertext {\Bigfonts \bf Divergences in the Casimir energy} \vskip3truept
\vskip 20truept
\centertext
{J.S.Dowker\footnote{dowker@a13.ph.man.ac.uk}} \vskip 7truept \centertext{\it
Department of Theoretical Physics,\\ The University of Manchester,
Manchester, England} \vskip10truept \vskip10truept \vskip7truept \vskip
20truept \centertext {Abstract} \vskip10truept
\begin{narrow}
The divergence found by Nesterenko, Lambiase and Scarpetta is attributed to
the existence of edges in the considered manifold.
\end{narrow}
\vskip 5truept
\vskip 60truept
\vfil
\end{title}
\pagenum=0
\section{\bf Introduction}

In a recent work, [\pref{NLS}], Nesterenko, Lambiase and Scarpetta encounter
extra divergences in the Casimir energy of a semi-circular infinite
cylinder. The present letter makes some brief comments on this
calculation.
\section{\bf Divergences in the presence of boundaries}

A general discussion, and some examples, can be found in
Dowker and Kennedy, [\pref{DandK}] sections 5 and 6 and some further 
remarks in Dowker and Banach [\pref{DandB}]. Let us work in $D$ spatial
dimensions, and assume space-time is ultrastatic $T\times\man$.
In the approach of Dowker and Critchley, [\pref{DandC}], the one-loop
effective Lagrangian for scalar fields is
$$
L^{(1)}=-{i\over2}\lim_{s\to1}{\tr_{\!D}
[\bze(s-1)]\over s-1}=
-{i\over2}\lim_{s\to1}\bigg({\tr_{\!D}[\bze(0)]\over s-1}+
\tr_{\!D}[\bze'(0)]\bigg)
\eql{efflag}$$
in terms of the space-time \zf, or, better, $\ze$-operator, $\bze(s)$. The
square brackets indicate that a time coincidence limit has been taken and the 
trace
can be taken as an integration over $\man$ of the space diagonal elements.
A scale dependent term could be added, but the pole is sufficient
for now. It can be shown, [\pref{DandK}], that the total vacuum energy
(the Casimir energy), $E_\man$, equals $-L^{(1)}$.

To connect $E$ with the \zf\ on the
spatial section, $\man$, we relate this latter \zf\ to the spacetime
\zf\ by [\pref{DandK}]
$$
[\bze(s)]={i\over(4\pi)^{1/2}}{\Ga(s-1/2)\over\Ga(s)}\,\bze_\man(s-1/2).
\eql{zfrel}$$

It can be seen from (\peq{efflag}) and (\peq{zfrel}) that if
$\ze_\man(s)$, $\equiv \tr_{\!D}\bze_\man(s)$, has no
pole at $s=-1/2$ then $E_\man$ is given by the finite expression
$$
E_\man={1\over2}\,\ze_\man(-1/2)
$$
used in [\pref{DandK,Gibb,DandB}] and many other places since.

We prefer this derivation of this result, rather than the
more usual one of simply regularising the divergent eigenvalue sum
form of the Casimir energy, since it arises via the effective
action, a scalar, invariant quantity whose
renormalisation (which we do not attempt) is more transparent
when divergences appear, as they will, in general. In
this case one finds the expression
$$
E_\man={1\over2}\,{\rm Pf}\ze_\man(-1/2)
+\lim_{s\to1}{C_{(D+1)/2}(\man)\over s-1}
\eql{cas}$$
the pole residue being given in terms of the $\man$ heat-kernel
expansion coefficient $C_{(D+1)/2}(\man)$. (For a conformally invariant theory,
this would be the conformal anomaly.)

For simplicity, let us now assume that $\man$ is flat. The coefficients
$C_n$ are then composed of just {\it boundary} parts, which we write
$B_n(\pa\man)$, the point now being that {\it if}  one combines the
Casimir energies for the inside, $\man$, and the outside, $\man^*$, of
the bounding surface, since $\pa\man=-\pa\man^*$ the divergent pole
terms in $E$ and $E^*$ from (\peq{cas}) will either cancel or
double according to the property, used in [\pref{DandK}],
$$
B_n(-\pa\man)=(-1)^{2n+1}\,B_n(\pa\man)
\eql{reflsym}$$
depending on whether $D$ is even or odd. The explicit
evaluations on the ball by Cognola, Elizalde and Kirsten, [\pref{CEK}],
illustrate this very nicely. It is to this particular calculation that the
authors of Ref [\pref{NLS}] refer and which they essentially repeat for the
semi-circular cylinder. In this geometry extra divergences do occur, 
even for the
outside--inside combination, because, as already pointed out in [\pref{DandK}],
the reflection symmetry, (\peq{reflsym}), no longer holds if the
domain $\man$ has edges or corners, or other singularities.

To the author's knowledge, a general treatment of this situation
does not yet exist but to proceed a little further we consider
in more detail the particular arrangement in [\pref{NLS}].
It is shown in [\pref{DandK}] section 6 that,
for a four-dimensional space-time, the vacuum energy per unit length of
an infinite waveguide of uniform cross-section, ${\cal D}$, is
$$
E_\cad={1\over8\pi}\bigg(\lim_{s\to1}{\ze_{\cal D}(-1)\over s-1}
+\ze_{\cal D}(-1)+\ze_{\cal D}'(-1)\bigg).
\eql{wgen}$$

The pole residue is proportional to the heat-kernel coefficient
$C_2({\cal D})$ which is, as has been stated, all boundary part
$B_2(\pa{\cal D})$, satisfying (\peq{reflsym}) if the boundary is smooth, which
is the most commonly referred to case. As stated, this pole will therefore
cancel on combining the outside and inside expressions. The finite term,
$\ze_{\cal D}(-1)$, whose coefficient is actually undetermined because of scale
ambiguities, also cancels leaving just, [\pref{DandK}],
$$E=E_\cad+E_{\cad^*}-E_{\cad\cup\cad^*}=
{1\over8\pi}\big(\ze'_\cad(-1)+\ze'_{\cad^*}(-1)-\ze'_{\cad\cup\cad^*}
(-1)\big)$$
as the finite Casimir energy associated with the existence of the 
(infinitely thin) surface $\pa\cad$ in the region ${\cad\cup\cad^*}$.

If the boundary is not smooth (say it is piecewise smooth) then we should
enumerate the pieces, $\pa{\cal D}_i$, which meet in the {\it
intersections} ${\cal I}_{ij}$ (here simply points) and one should properly
say that $C_2$ takes contributions from the $\pa{\cal D}_i$ and the ${\cal
I}_{ij}$ on the same footing. It is these latter contributions that violate
the reflection symmetry (\peq{reflsym}) and are responsible for any extra
divergences found in [\pref{NLS}]. They typically involve the {\it squares}
of the extrinsic curvatures.

As the particular example under consideration, the heat-kernel expansion on a
semi-disc (a
hemi-one ball), and indeed on an arbitrary sector, is easily worked out
either from the eigenmodes,  {\it \`a la} Stewartson and Waechter, or Moss
[\pref{Moss}], or from images (when they apply) or from \zfs\ or from a
combination of techniques. In the case of the semicircular boundary, the
extra term, over and above the `circular' one, arises from the image
contribution.

For the record the semi-circle expansion is
$$
K(t)\sim {|{\cal D}|\over8\pi t}-{5\over16}{|\pa{\cal D}|\over(\pi t)^{1/2}}+
{5\over24}+{(\pi t)^{1/2}\over16}\bigg({1\over16}+{1\over\pi}\bigg)+
{347\over10080}\,t+\,\,.
$$

A tolerably comprehensive analysis of the general $C_2$ coefficient
in the presence of boundary discontinuities in arbitrary dimensions
has been attempted in Ref [\pref{AandD}]. If it is required
to eliminate the divergences by any form of renormalisation
then it is necessary to motivate bare quantities of the general
form given in [\pref{AandD}]. About this we have nothing to say. Without
physical justification for studying such singular manifolds (\eg
brane worlds), perhaps one
should not strive too hard in this direction especially since
in the generic case, where $C_{(D+1)/2}$ takes contributions from submanifolds
of codimension up to the dimension of the manifold, dimensional
arguments alone show that pole cancellation is impossible.
\section{\bf Acknowledgment}

I thank Klaus Kirsten for helpful discussions. This work was performed with
support from EPSRC grant no GR/M08714.
\section{\bf References}
\vskip 5truept
\begin{putreferences}
\ref{APS}{Atiyah,M.F., V.K.Patodi and I.M.Singer: Spectral asymmetry and
Riemannian geometry \mpcps{77}{75}{43}.}
\ref{AandD}{Apps,J.S. and Dowker,J.S. \cqg{15}{98}{1121}.}
\ref{AandT}{Awada,M.A. and D.J.Toms: Induced gravitational and gauge-field
actions from quantised matter fields in non-abelian Kaluza-Klein thory
\np{245}{84}{161}.}
\ref{BandI}{Baacke,J. and Y.Igarishi: Casimir energy of confined massive
quarks \prD{27}{83}{460}.}
\ref{Barnesa}{Barnes,E.W.: On the Theory of the multiple Gamma function
{\it Trans. Camb. Phil. Soc.} {\bf 19} (1903) 374.}
\ref{Barnesb}{Barnes,E.W.: On the asymptotic expansion of integral
functions of multiple linear sequence, {\it Trans. Camb. Phil.
Soc.} {\bf 19} (1903) 426.}
\ref{Barv}{Barvinsky,A.O. Yu.A.Kamenshchik and I.P.Karmazin: One-loop
quantum cosmology \aop {219}{92}{201}.}
\ref{BandM}{Beers,B.L. and Millman, R.S. :The spectra of the
Laplace-Beltrami
operator on compact, semisimple Lie groups. \ajm{99}{1975}{801-807}.}
\ref{BandH}{Bender,C.M. and P.Hays: Zero point energy of fields in a
confined volume \prD{14}{76}{2622}.}
\ref{BBG}{Bla\v zi\' c,N., Bokan,N. and Gilkey,P.B.: Spectral geometry of the
form valued Laplacian for manifolds with boundary \ijpam{23}{92}{103-120}}
\ref{BEK}{Bordag,M., E.Elizalde and K.Kirsten: { Heat kernel
coefficients of the Laplace operator on the D-dimensional ball},
\jmp{37}{96}{895}.}
\ref{BGKE}{Bordag,M., B.Geyer, K.Kirsten and E.Elizalde,: { Zeta function
determinant of the Laplace operator on the D-dimensional ball},
\cmp{179}{96}{215}.}
\ref{BKD}{Bordag,M., K.Kirsten,K. and Dowker,J.S.: Heat kernels and
functional determinants on the generalized cone \cmp{182}{96}{371}.}
\ref{Branson}{Branson,T.P.: Conformally covariant equations on differential
forms \cpde{7}{82}{393-431}.}
\ref{BandG2}{Branson,T.P. and P.B.Gilkey {\it Comm. Partial Diff. Eqns.}
{\bf 15} (1990) 245.}
\ref{BGV}{Branson,T.P., P.B.Gilkey and D.V.Vassilevich {\it The Asymptotics
of the Laplacian on a manifold with boundary} II, hep-th/9504029.}
\ref{BCZ1}{Bytsenko,A.A, Cognola,G. and Zerbini, S. : Quantum fields in
hyperbolic space-times with finite spatial volume, hep-th/9605209.}
\ref{BCZ2}{Bytsenko,A.A, Cognola,G. and Zerbini, S. : Determinant of
Laplacian on a non-compact 3-dimensional hyperbolic manifold with finite
volume, hep-th /9608089.}
\ref{CandH2}{Camporesi,R. and Higuchi, A.: Plancherel measure for $p$-forms
in real hyperbolic space, \jgp{15}{94}{57-94}.}
\ref{CandH}{Camporesi,R. and A.Higuchi {\it On the eigenfunctions of the
Dirac operator on spheres and real hyperbolic spaces}, gr-qc/9505009.}
\ref{ChandD}{Chang, Peter and J.S.Dowker :Vacuum energy on orbifold factors
of spheres, \np{395}{93}{407}.}
\ref{cheeg1}{Cheeger, J.: Spectral Geometry of Singular Riemannian Spaces.
\jdg {18}{83}{575}.}
\ref{cheeg2}{Cheeger,J.: Hodge theory of complex cones {\it Ast\'erisque}
{\bf 101-102}(1983) 118-134}
\ref{Chou}{Chou,A.W.: The Dirac operator on spaces with conical
singularities and positive scalar curvature, \tams{289}{85}{1-40}.}
\ref{CandT}{Copeland,E. and Toms,D.J.: Quantized antisymmetric
tensor fields and self-consistent dimensional reduction
in higher-dimensional spacetimes, \break\np{255}{85}{201}}
\ref{DandH}{D'Eath,P.D. and J.J.Halliwell: Fermions in quantum cosmology
\prD{35}{87}{1100}.}
\ref{cheeg3}{Cheeger,J.:Analytic torsion and the heat equation. \aom{109}
{79}{259-322}.}
\ref{CEK}{Cognola,G.,Elizalde, E. and Kirsten,K. `Casimir energies for
spherically symmetric cavities', hep-th/9906228.}
\ref{DandE}{D'Eath,P.D. and G.V.M.Esposito: Local boundary conditions for
Dirac operator and one-loop quantum cosmology \prD{43}{91}{3234}.}
\ref{DandE2}{D'Eath,P.D. and G.V.M.Esposito: Spectral boundary conditions
in one-loop quantum cosmology \prD{44}{91}{1713}.}
\ref{Dow1}{Dowker,J.S.: Effective action on spherical domains, \cmp{162}{94}
{633}.}
\ref{Dow8}{Dowker,J.S. {\it Robin conditions on the Euclidean ball}
MUTP/95/7; hep-th\break/9506042. {\it Class. Quant.Grav.} to be published.}
\ref{Dow9}{Dowker,J.S. {\it Oddball determinants} MUTP/95/12;
hep-th/9507096.}
\ref{Dow10}{Dowker,J.S. {\it Spin on the 4-ball},
hep-th/9508082, {\it Phys. Lett. B}, to be published.}
\ref{DandA2}{Dowker,J.S. and J.S.Apps, {\it Functional determinants on
certain domains}. To appear in the Proceedings of the 6th Moscow Quantum
Gravity Seminar held in Moscow, June 1995; hep-th/9506204.}
\ref{DandK}{Dowker,J.S. and Kennedy,G. \jpa{11}{78}{895}.}
\ref{DandB}{Dowker,J.S. and Banach,R. \jpa{11}{78}{2255}.}
\ref{DandC}{Dowker,J.S. and Critchley,R. \prD{13}{76}{2224}.}
\ref{DABK}{Dowker,J.S., Apps,J.S., Bordag,M. and Kirsten,K.: Spectral
invariants for the Dirac equation with various boundary conditions
{\it Class. Quant.Grav.} to be published, hep-th/9511060.}
\ref{EandR}{E.Elizalde and A.Romeo : An integral involving the
generalized zeta function, {\it International J. of Math. and
Phys.} {\bf13} (1994) 453.}
\ref{ELV2}{Elizalde, E., Lygren, M. and Vassilevich, D.V. : Zeta function
for the laplace operator acting on forms in a ball with gauge boundary
conditions. hep-th/9605026}
\ref{ELV1}{Elizalde, E., Lygren, M. and Vassilevich, D.V. : Antisymmetric
tensor fields on spheres: functional determinants and non-local
counterterms, \jmp{}{96}{} to appear. hep-th/ 9602113}
\ref{Kam2}{Esposito,G., A.Y.Kamenshchik, I.V.Mishakov and G.Pollifrone:
Gravitons in one-loop quantum cosmology \prD{50}{94}{6329};
\prD{52}{95}{3457}.}
\ref{Erdelyi}{A.Erdelyi,W.Magnus,F.Oberhettinger and F.G.Tricomi {\it
Higher Transcendental Functions} Vol.I McGraw-Hill, New York, 1953.}
\ref{Esposito}{Esposito,G.: { Quantum Gravity, Quantum Cosmology and
Lorentzian Geometries}, Lecture Notes in Physics, Monographs, Vol. m12,
Springer-Verlag, Berlin 1994.}
\ref{Esposito2}{Esposito,G. {\it Nonlocal properties in Euclidean Quantum
Gravity}. To appear in Proceedings of 3rd. Workshop on Quantum Field Theory
under the Influence of External Conditions, Leipzig, September 1995;
gr-qc/9508056.}
\ref{EKMP}{Esposito G, Kamenshchik Yu A, Mishakov I V and Pollifrone G.:
One-loop Amplitudes in Euclidean quantum gravity.
\prd {52}{96}{3457}.}
\ref{ETP}{Esposito,G., H.A.Morales-T\'ecotl and L.O.Pimentel {\it Essential
self-adjointness in one-loop quantum cosmology}, gr-qc/9510020.}
\ref{FORW}{Forgacs,P., L.O'Raifeartaigh and A.Wipf: Scattering theory, U(1)
anomaly and index theorems for compact and non-compact manifolds
\np{293}{87}{559}.}
\ref{GandM}{Gallot S. and Meyer,D. : Op\'erateur de coubure et Laplacian
des formes diff\'eren-\break tielles d'une vari\'et\'e riemannienne
\jmpa{54}{1975}{289}.}
\ref{Gibb}{Gibbons,G.W.\pl{60A}{77}{385}.}
\ref{Gilkey1}{Gilkey, P.B, Invariance theory, the heat equation and the
Atiyah-Singer index theorem, 2nd. Edn., CTC Press, Boca Raton 1995.}
\ref{Gilkey2}{Gilkey,P.B.:On the index of geometric operators for
Riemannian manifolds with boundary \aim{102}{93}{129}.}
\ref{Gilkey3}{Gilkey,P.B.: The boundary integrand in the formula for the
signature and Euler characteristic of a manifold with boundary
\aim{15}{75}{334}.}
\ref{Grubb}{Grubb,G. {\it Comm. Partial Diff. Eqns.} {\bf 17} (1992)
2031.}
\ref{GandS1}{Grubb,G. and R.T.Seeley \cras{317}{1993}{1124}; \invm{121}{95}
{481}.}
\ref{GandS}{G\"unther,P. and Schimming,R.:Curvature and spectrum of compact
Riemannian manifolds, \jdg{12}{77}{599-618}.}
\ref{IandT}{Ikeda,A. and Taniguchi,Y.:Spectra and eigenforms of the
Laplacian
on $S^n$ and $P^n(C)$. \ojm{15}{1978}{515-546}.}
\ref{IandK}{Iwasaki,I. and Katase,K. :On the spectra of Laplace operator
on $\La^*(S^n)$ \pja{55}{79}{141}.}
\ref{JandK}{Jaroszewicz,T. and P.S.Kurzepa: Polyakov spin factors and
Laplacians on homogeneous spaces \aop{213}{92}{135}.}
\ref{Kam}{Kamenshchik,Yu.A. and I.V.Mishakov: Fermions in one-loop quantum
cosmology \prD{47}{93}{1380}.}
\ref{KandM}{Kamenshchik,Yu.A. and I.V.Mishakov: Zeta function technique for
quantum cosmology {\it Int. J. Mod. Phys.} {\bf A7} (1992) 3265.}
\ref{KandC}{Kirsten,K. and Cognola.G,: { Heat-kernel coefficients and
functional determinants for higher spin fields on the ball} \cqg{13}{96}
{633-644}.}
\ref{Levitin}{Levitin,M.: { Dirichlet and Neumann invariants for Euclidean
balls}, {\it Diff. Geom. and its Appl.}, to be published.}
\ref{Luck}{Luckock,H.C.: Mixed boundary conditions in quantum field theory
\jmp{32}{91}{1755}.}
\ref{MandL}{Luckock,H.C. and Moss,I.G,: The quantum geometry of random
surfaces and spinning strings \cqg{6}{89}{1993}.}
\ref{Ma}{Ma,Z.Q.: Axial anomaly and index theorem for a two-dimensional
disc
with boundary \jpa{19}{86}{L317}.}
\ref{Mcav}{McAvity,D.M.: Heat-kernel asymptotics for mixed boundary
conditions \cqg{9}{92}{1983}.}
\ref{MandV}{Marachevsky,V.N. and D.V.Vassilevich {\it Diffeomorphism
invariant eigenvalue \break problem for metric perturbations in a bounded
region}, SPbU-IP-95, \break gr-qc/9509051.}
\ref{Milton}{Milton,K.A.: Zero point energy of confined fermions
\prD{22}{80}{1444}.}
\ref{MandS}{Mishchenko,A.V. and Yu.A.Sitenko: Spectral boundary conditions
and index theorem for two-dimensional manifolds with boundary
\aop{218}{92}{199}.}
\ref{Moss}{Moss,I.G.
\cqg{6}{89}{759}.}
\ref{MandP}{Moss,I.G. and S.J.Poletti: Conformal anomaly on an Einstein space
with boundary \pl{B333}{94}{326}.}
\ref{MandP2}{Moss,I.G. and S.J.Poletti \np{341}{90}{155}.}
\ref{NLS}{Nesterenko,V.V., Lambiase,G. and Scarpetta,G. `Casimir Energy
of a Semi-Circular Infinite Cylinder', hep-th/0005257.}
\ref{NandOC}{Nash, C. and O'Connor,D.J.: Determinants of Laplacians, the
Ray-Singer torsion on lens spaces and the Riemann zeta function
\jmp{36}{95}{1462}.}
\ref{NandS}{Niemi,A.J. and G.W.Semenoff: Index theorem on open infinite
manifolds \np {269}{86}{131}.}
\ref{NandT}{Ninomiya,M. and C.I.Tan: Axial anomaly and index thorem for
manifolds with boundary \np{245}{85}{199}.}
\ref{norlund2}{N\"orlund~N. E.:M\'emoire sur les polynomes de Bernoulli.
\am {4}{21} {1922}.}
\ref{Poletti}{Poletti,S.J. \pl{B249}{90}{355}.}
\ref{RandT}{Russell,I.H. and Toms D.J.: Vacuum energy for massive forms
in $R^m\times S^N$, \cqg{4}{86}{1357}.}
\ref{RandS}{R\"omer,H. and P.B.Schroer \pl{21}{77}{182}.}
\ref{Trautman}{Trautman,A.: Spinors and Dirac operators on hypersurfaces
\jmp{33}{92}{4011}.}
\ref{Vass}{Vassilevich,D.V.{Vector fields on a disk with mixed
boundary conditions} gr-qc /9404052.}
\ref{Voros}{Voros,A.:
Spectral functions, special functions and the Selberg zeta function.
\cmp{110}{87}439.}
\ref{Ray}{Ray,D.B.: Reidemeister torsion and the Laplacian on lens
spaces \aim{4}{70}{109}.}
\ref{McandO}{McAvity,D.M. and Osborne,H. Asymptotic expansion of the heat 
kernel for generalised boundary conditions \cqg{8}{91}{1445}.}
\ref{AandE}{Avramidi,I. and Esposito,G.}
\ref{MandS}{Moss,I.G. and Silva P.J. Invariant boundary conditions for
gauge theories gr-qc/9610023.}
\ref{barv}{Barvinsky,A.O.\pl{195B}{87}{344}.}
\end{putreferences}

\bye